\documentclass{aastex61}
\usepackage{natbib}
\usepackage{amsmath}
\usepackage{changepage}

\setcitestyle{notesep={, }, aysep={; }}

\shorttitle{Broadband Photometric Reverberation Mapping of Quasars}
\shortauthors{Zhang et al.}

\begin{document}

\title{Broadband Photometric Reverberation Mapping Analysis on SDSS-RM and Stripe 82 Quasars}

\correspondingauthor{Xue-Bing Wu}
\email{wuxb@pku.edu.cn}

\author{Haowen Zhang}
\affiliation{School of Physics, Peking University, Beijing 100871, China}
\affiliation{Kavli Institute of Astronomy and Astrophysics, Peking University, Beijing 100871, China}

\author{Qian Yang}
\affiliation{Department of Astronomy, School of Physics, Peking University, Beijing 100871, China}
\affiliation{Kavli Institute of Astronomy and Astrophysics, Peking University, Beijing 100871, China}

\author{Xue-Bing Wu}
\affiliation{Department of Astronomy, School of Physics, Peking University, Beijing 100871, China}
\affiliation{Kavli Institute of Astronomy and Astrophysics, Peking University, Beijing 100871, China}

\begin{abstract}
We modified the broadband photometric reverberation mapping (PRM) code, JAVELIN, and tested the availability to get broad line region (BLR) time delays that are consistent with the spectroscopic reverberation mapping (SRM) project SDSS-RM. The broadband light curves of SDSS-RM quasars produced by convolution with the system transmission curves were used in the test. We found that under similar sampling conditions (evenly and frequently sampled), the key factor determining whether the broadband PRM code can yield lags consistent with the SRM project is the flux ratio of the broad emission line to the reference continuum, which is in line with the previous findings. We further found a critical line-to-continuum flux ratio, about 6\%, above which the mean of the ratios between the lags from PRM and SRM becomes closer to unity, and the scatter is pronouncedly reduced. We also tested our code on a subset of SDSS Stripe 82 quasars, and found that our program tends to give biased lag estimations due to the observation gaps when the R-L relation prior in Markov Chain Monte Carlo (MCMC) is discarded. The performance of the damped random walk (DRW) model and the power-law (PL) structure function model on broadband PRM were compared. We found that given both SDSS-RM-like or Stripe 82-like light curves, the DRW model performs better in carrying out broadband PRM than the PL model.  
\end{abstract}


\section{Introduction} 
\label{sec:introduction}


The measurement of the masses of supermassive black holes (SMBH) is an important issue in the study of the growth and interaction of SMBHs and their host galaxies. Reverberation mapping (RM) \citep{Blandford:1982il,Gaskell:1986bb,Anonymous:N0Liuza0}, which utilizes the variability of active galactic nuclei (AGN), has been an established way in this field. The physical picture of reverberation mapping is straightforward: The broad line region (BLR) of an AGN, which is an unresolved region made up of photoionized gas with high density, is excited by the continuum photons from the central engine, and radiates line emissions. Because of the high velocity motions of the re-emitting materials, the line profiles are significantly broadened as a result of Doppler effect. Due to the non-negligible spatial separation between the central accretion disk and the BLR and due to the geometry of the BLR, the variability pattern of the BLR radiation is delayed and modulated compared with that of the continuum which originates from the central region around the black hole. Quantitatively, the radiation flux from the BLR can be modeled using the following equation:

\begin{equation}
	f_{BLR}(t) = \int d\tau\ \Psi(\tau) f_{cont}(t - \tau),
\end{equation}
where $\tau$ is the light traveling delay from the central engine to the BLR, and $\Psi(\tau)$ is the so-called transfer function determined by the geometry of the BLR. \citet{Blandford:1982il} gave detailed discussions on the specific forms of transfer functions under various geometry and accretion disk models, and presented the calculation of the variability of the emission lines for some simple models.

By measuring this delay we can effectively estimate the physical size of the BLR. \citet{Gaskell:1986bb} measured the BLR sizes of 3 Seyfert I galaxies under the assumption that the response of the BLR to the continuum is of a power-law form. If we further assume that the motion of the BLR clouds is Keplerian, which has been demonstrated by a special example of NGC5548 in \citet{Peterson:1999cf}, we have the following equation to estimate the mass of the central black hole $M_{BH}$ :

\begin{equation}
	M_{BH} = f \frac{\Delta V^2 c\ \tau_{delay}}{G},
\end{equation}
where $\Delta V$ is the typical velocity of the BLR clouds, which can be estimated from the line width of broad emission lines, $c$ is the speed of light, $G$ is the gravitational constant, and $f$ is a scaling factor depending on the specific geometry of the BLR \citep{Peterson:1999cf}. With this method, \citet{Peterson:1999cf} measured the mass of the SMBH in the Seyfert I galaxy NGC5548 using the time lags and line widths of several broad emission lines, in which the scaling factor $f$ is assumed to be unity and the systematic error is neglected. On the mean value of $f$ there are still some active discussions. \citep[e.g.,][]{Woo:2010ie,Graham:2011hc}

Previously the bulk of reverberation mapping work was done spectroscopically. In this way multi-epoch spectra of the target AGNs/quasars were taken and fitted to extract the variability of the broad emission lines. A prevalent method to determine the lag between emission line and continuum light curves is cross-correlation function (CCF) analysis and its variants, including interpolation CCF (ICCF) \citep{Gaskell:1986bb,White:1994gj,Peterson:1998dx,Anonymous:N0Liuza0}, discrete correlation function (DCF) \citep{Edelson:1988bz} and Z-transformed DCF (ZDCF) \citep{Alexander:1997kw,Anonymous:N0Liuza0}. All these CCF methods share a key assumption that the light curve containing contribution from the broad-line emission must be dominated by it, otherwise the peak at the position of a non-zero lag would be overwhelmed by the auto-correlation peak of the continuum at zero lags. Thus generally CCF-like methods only apply to spectroscopic reverberation mapping, sometimes also to broadband+narrowband photometric RM \citep[e.g.,][]{Haas:2011fy,PozoNunez:2013fm,Anonymous:RGAl7jzU}, in which, however, the contribution from the continuum to the narrow band light curve must be subtracted carefully. There are also some works applying CCF-like methods directly to broadband photometric data \citep[e.g.,][]{Anonymous:nxpdmZmg}, but due to the fact that the typical relative contribution of a broad emission line to the whole broadband is of the order of a few percent, this method cannot be expected to perform well on most of the quasars whose variability information comes only from their broadband photometric data. Another problem of CCF-like methods is that the accuracy of the recovered time lags is highly affected by the sampling quality of the light curves. \citet{White:1994gj} demonstrated that the recovered time lags from ICCF, where the light curves are interpolated linearly between the real data points, becomes unreliable when the mean sampling interval gets larger than $\sim$14 days. On the other hand, they argued that an adequate number of data points, which is generally equivalent to frequent enough sampling, is crucial for DCF, in which the lag is only calculated using the real observed data. However, \citet{2013ApJ...769..124C} generalized the CCF method into multivariate correlation functions (MCF) and suggested the availability of broadband photometric reverberation mapping on quasar light curves of high qualities.

Some previous works \citep{Kelly:2009fx,Zu:2011ka} have demonstrated that the variability of quasars can be well modeled as damped random walk (DRW), in which only two parameters, the random walk amplitude $\sigma$ and the relaxation timescale $\tau$, are used to describe the optical variability of quasars. By modeling the light curves in this way, one is able to predict the magnitude or flux on an arbitrary time in a more natural way than using ICCF, where the light curves are linearly interpolated between the observed points. \citet{2010ApJ...721.1014M} utilized the broadband photometric data of the quasars in SDSS Stripe 82 \citep{Abazajian:2009ef}, and found several scaling relationships between the DRW parameters and other intrinsic properties of quasars. \citet{Anonymous:H3eiB3AM} developed a new method to carry out quasar selection based on the DRW model. There have also been some generalizations on the DRW model, where the stochastic process reduces to DRW in certain special cases \citep[e.g.,][]{Zu:2012cg,2011ApJ...730..139P}. On the other hand, there also have been some works focused on using the structure function method to study the variabilities of quasars. \citet{Collier:2001ey} measured the optical structure functions of 12 AGNs. \citet{Anonymous:NTBTE_Lk} calculated the ensemble structure function using the calibrated historical photometric data of 401 quasars, and found no turnover at the timescale of $\sim$40 years. Using SDSS Stripe 82 photometric data, \citet{2010ApJ...714.1194S} showed that when modeled as a single power-law function of time, the structure function parameters, i.e. the amplitude $A$ and the power-law index $\gamma$ of quasar light curves populate a distinct region in the parameter space from those of non-variable or other variable sources. \citet{Anonymous:DtBp8yhK} developed an unbiased way of measuring the decorrelation timescales which utilizes the structure function instead of fitting the light curve to a DRW process. 

Based on the DRW model, some works have tried utilizing pure broadband photometric data to do the reverberation mapping \citep[where the latter also used the PL model]{2010ascl.soft10007Z,Zu:2011ka,2016ApJ...819..122Z,Hernitschek:2015iw}, which is of much larger data volume and easier to get access to. However, \citet{Hernitschek:2015iw} imposed an additional prior probability term in their Markov Chain Monte Carlo (MCMC) code, which leads to a preference to lag values closer to the R-L relation obtained by \citet{Anonymous:N0Liuza0}. We found that this term tends to overwhelm the other term decided by the damped random walk in the posterior probability, which leaves the good match between their results and the R-L relation somewhat dubious. Thus trying to obtain time lag estimations that are independent of the R-L relation is one of the motivations of this work. Also \citet{Hernitschek:2015iw} did not get access to any photometric data that are of high and even sampling cadences like what were produced in this work. 

In this work, we modified and extended the JAVELIN code (made by \citet{2010ascl.soft10007Z}) in order to explore the availability of pure broadband photometric reverberation mapping on two datasets with considerably different sampling qualities: SDSS-RM quasars and SDSS Stripe 82 quasars. We also tried to discard the prior term which prefers lag values closer to the R-L relation, in order to see to what extent the DRW (or PL) model can perform given highly sparse and unevenly sampled data.

The paper is organized as follows. Section 2 covers the methodology utilized in this work, including a brief introduction to broadband photometric reverberation mapping and the modification made based on the original JAVELIN code. Also we describe the process in which the original SDSS-RM spectra were converted into broadband optical light curves. In section 3 we give an overview of our data, and describe some of their key qualities in this work. In Section 4 we present the results of the code using both datasets, and the discussions are given in Section 5. In section 6 we summarize the paper.

\section{Methodology}

\subsection{General Idea of Broadband Photometric Reverberation Mapping} 
\label{sub:general_idea_of_broadband_photometric_reverberation_mapping}

The general idea of broadband photometric reverberation mapping is simple: to determine the variability of broad emission lines from broadband photometric data as accurately and precisely as possible, and then compare it with that of the continuum in order to determine the time lag.

To do this, the JAVELIN code, whether the original version or the modified one used in this work, essentially models the underlying stochastic process as DRW process (and power-law model in the modified version), calculates the theoretical light curves of both of the broadbands given a certain set of parameters, and then find the parameter set that optimizes the match between the calculated and observed light curves using the  MCMC algorithm.

The complete formalism can be found in \citet{Anonymous:FfspGr5o}, \citet{2010ascl.soft10007Z} and \citet{Hernitschek:2015iw}, and we only briefly present the key formalisms here.

Our method assumes that the data vector $\mathbf{y}$ consists of three parts: underlying true signal $\mathbf{s}$, noise $\mathbf{n}$ and the general trend term $L\mathbf{q}$, i.e. $\mathbf{y} = \mathbf{s} + \mathbf{n} + L\mathbf{q}$, where $L$ is a $\mathbf m\times \mathbf n$ matrix if we are using $\mathbf{m}$ light curves, each having $\mathbf{n}$ data points, and $\mathbf{q}$ is a set of linear coefficients to be fitted. The overall covariance matrix can be decomposed into two parts: $C = S + N$, where $S$ and $N$ are the covariance matrices of the signal and noise components, respectively.

Based on this decomposition, the posterior probability that the observed data occurs under a given set of parameters (which determines the form of the overall covariance matrix $C$) is \citep{Zu:2011ka}:

\begin{equation}
P\left(\mathbf{y}|\mathbf{p}\right) \propto \mathcal{L} \equiv |S+N|^{-1/2}|L^T C^{-1} L|^{-1/2}\exp\left(-\frac{\mathbf{y}^T C_{\perp}^{-1} \mathbf{y}}{2}\right),
\end{equation}
where $\mathcal{L}$ is the likelihood function, and
\begin{equation}
C_{\perp}^{-1} = C^{-1} - C^{-1} L C_q L^T C^{-1},\ C_q = \left(L^T C^{-1} L\right)^{-1},
\end{equation}

When we model the quasar's optical variability as a damped random walk process, the covariance matrix $C$ takes the following form:

\begin{equation}
	C_{ij} = \mathrm{Cov} \langle f_{cont}\left(t_i\right), f_{cont}\left(t_j\right) \rangle = \sigma^2 \exp\left(-\frac{|t_i - t_j|}{\tau}\right),
\end{equation}
where $\sigma$ is the DRW amplitude of variability and $\tau$ is the relaxation timescale.

Combining Eq. (5) and Eq. (1) one can compute the covariance term between a pure broad emission line flux and the continuum:

\begin{equation}
\begin{split}
C_{ij} = \mathrm{Cov} \langle f_{line}\left(t_i\right), f_{cont}\left(t_j\right) \rangle = & \mathrm{Cov} \langle \int f_{cont}\left(t_i - t'\right) \Psi\left(t'\right)dt', f_{cont}\left(t_j\right)\rangle = \\
\int \Psi\left(t'\right) \mathrm{Cov} \langle f_{cont}\left(t_i - t'\right), f_{cont}\left(t_j\right)\rangle dt' = & \sigma^2 \int \Psi\left(t'\right)\exp\left(-\frac{t_i - t_j - t'}{\tau}\right) dt'.
\end{split}
\end{equation}

In practice, aside from the pure continuum band fluxes, what are dumped into the calculations are the broadband fluxes which encloses a broad emission line along with continuum (line+continuum band) contributions:

\begin{equation}
	f_{line+cont}\left(t_i\right) = f_{line}^{line+cont}\left(t_i\right) + f_{cont}^{line+cont}\left(t_i\right) = s_{line}\int \Psi\left(\boldsymbol{\tau_{delay}}\right)f^{cont}_{cont}\left(t_i - \boldsymbol{\tau_{delay}}\right)d\boldsymbol{\tau_{delay}} + \alpha f^{cont}_{cont}\left(t_i\right),
\end{equation}
where the superscripts denote the bands which the fluxes belong to (pure continuum band or line+continuum band). So in fact four more parameters are needed to determine the whole covariance matrix: the first one, of course, is the lag between the broad line and the continuum, along with the width of the top-hat transfer function $w$. The other two parameters are the flux ratios of emission line and the continuum in the line+continuum band to the pure continuum band, namely $s_{line}$ and $\alpha$, which determines the magnitude of the corresponding covariance terms. In short, the parameter set to be determined by the MCMC (under the DRW model) is ($\sigma$, $\tau$, $\tau_{delay}$, $w$, $s_{line}$, $\alpha$).

For the full formalism of all the covariance matrix entries under the DRW model and the top-hat transfer function assumption, we refer the readers to the Appendix of \citet{Zu:2011ka}.

In order to constrain the program better, we multiply the likelihood by a series of prior probabilities and the final product is used as the posterior function in MCMC. By maximizing the posterior function, the JAVELIN code gives the optimal parameter set for each object. Among these parameters, the one of the most interest to us is the lag between the broad emission lines and the continuum emissions.


\newpage
\subsection{Modification on the Code}
\subsubsection{Addition of Power-Law Model}
We added another descriptive model for quasar variability in which the structure function of a light curve is modeled as a power-law function of the time interval between two observations:

\begin{equation}
\mathrm{SF}(\Delta t) = A^2\left(\frac{\Delta t}{1\ \mathrm{yr} }\right)^{\gamma},
\end{equation}
where $A$ is the amplitude of the structure function, and $\gamma$ is the power-law index. Note that the model is only valid for the segment of the structure function where $\mathrm{SF}(\Delta t)$ increases monotonically with $\Delta t$, i.e. before it flattens at a longer timescale \citep{Kelly:2009fx,2010ApJ...721.1014M,Anonymous:H3eiB3AM}. 

Given this form of structure function, we can calculate the covariance term between the two pure continuum bands \citep{Hernitschek:2015iw}:

\begin{equation}
\mathrm{Cov} \langle f_{cont}\left(t_i\right), f_{cont}\left(t_j\right) \rangle = A^2 \left[\left(\frac{t_{\infty}}{1\ \mathrm{yr} }\right)^{\gamma} - \frac{1}{2} \left(\frac{|t_i-t_j|}{1\ \mathrm{yr} }\right)^{\gamma}\right].
\end{equation}

In this case the parameter set to be determined in MCMC is ($A$, $\gamma$, $\tau_{delay}$, $w$, $s_{line}$, $\alpha$). Other relevant covariance terms used in the code  under the power-law model are listed in the Appendix.

\subsubsection{Various Priors in the Posterior Propability Calculation}
Due to the mathematical nature of the algorithm, it sometimes produces dubious lags that are mathematically optimal but apparently unphysical. Based on this fact we impose some priors in addition to those implemented in the original JAVELIN code to try to rule them out as much as possible. In the following we list all the new priors we have imposed and their motivations:

1. We rule out all the negative lags or those larger than 10 times of the corresponding SRM results from \citet{Shen:2015fn} for the SDSS-RM quasars, by imposing a negative infinite probability on these sampled points in MCMC. A similar cutoff is done for SDSS Stripe 82 quasars, but since no SRM results are available for them, we use the lag estimations from the R-L empirical relation from \citet{Bentz:2013hm} instead. The motivation is straightforward: Firstly the negative lags are obviously unphysical, and the most of the PRM lags should not deviate from the empirical relationship obtained from SRM too much, or they may be presumably dubious. Note that in a previous work, \citet{Hernitschek:2015iw} imposed a $(0.25\tau_{R-L},4\tau_{R-L})$ cutoff in MCMC (where $\tau_{R-L}$ is the time lag estimation from the empirical R-L relation) along with a prior which peaks at the lag value obtained from R-L relation. Compared to their work we enlarged the cutoff window and discarded the prior term to test whether our algorithm gives reasonable results only because of the relatively narrow cutoff window and the selection effect induced by the prior term.

2. We fix the line-to-continuum flux ratio when calculating the lag. The motivation for this implementation is rather empirical: in many cases if we leave the ratio a free parameter, the code ends up yielding unphysical ratio values (sometimes even much larger than unity), which may affect the performance in the lag calculation. Although the line-to-continuum flux ratios are calculated from the RMS spectra of each quasar, their values at each epoch do not show significant deviation. Thus using the line-to-continuum flux ratios from the RMS spectra is a reasonable approximation.

3. We constrain the contribution of the continuum to the continuum+line band in the following way: We first calculate the mean of the line+continuum band and pure continuum band fluxes, $\langle f_{line+cont} \rangle$ and $\langle f_{cont} \rangle$, respectively, take their ratio, $\alpha_0 = \langle f_{line+cont} \rangle/\langle f_{cont} \rangle$ as an estimation of the ratio between the fluxes of the two bands at each time step, and rule out all the sampled points with $\alpha > 1.2\alpha_0$. The motivation is pretty much the same as in the second prior, but with the additional fact that for most of the quasars the continuum dominates the line+continuum band, which enables us to estimate the flux ratio between the two bands directly using the broadband photometric data.

4. Last but not least, we emphasize that the prior giving larger probability to lag values matching the R-L relation better, which was implemented by \citet{Hernitschek:2015iw}, is not used in our work. The reason is that we found this term tends to dominate the posterior probability, making the time lag measurement much more reliant on the empirical R-L relation. What's more, this dependency on the R-L relation may in fact hamper the discovery of outliers to the relation, which is of more physical interest.

\subsubsection{The Determination of Lag Values from MCMC Results of JAVELIN}
In the original JAVELIN, the highest posterior density (HPD, namely the 16\%-84\% percentile range of each parameter determined by MCMC) intervals and the best fit parameters where the posterior in logarithm is maximized are given. However we find neither of them produced in a single MCMC run yields satisfactory (i.e. consistent with the SRM values) estimates of the lags. Thus we use the following way instead to get our estimations:\\

1. We run 200 MCMC runs for each tested quasar and collect their best fit parameters.

2. We perform a kernel density estimation (KDE) on the yielded best fit lag distribution, using a Gaussian kernel with a bandwidth determined by 3-fold cross-validation, which is implemented in Sci-kit Learn \citep{Pedregosa:2011tv}.

3. We follow what previous SRM projects did on the CCFs: We define $\tau_{peak}$ as the lag value where the KDE value maximizes, $\tau_{cent}$ as the centroid value in the lag intervals where the KDE value is over 80\% of the maximum \citep[e.g.,][]{Peterson:1998dx}. When calculating the corresponding $\sigma_{\tau}$, due to the limited computational resources we were not able to calculate them using ``flux randomization/random subset selection" (``FR/RSS") method \citep{Peterson:1998dx,Shen:2015fn}. Instead they are determined by the FWHM of the KDE profile: $\sigma_{\tau} = \rm FWHM/\left(2\sqrt{ln2}\right)$.


Finally we point out that in all the calculations we use the light curves in unit of flux rather than magnitude. We adopt this method since the whole formalism of JAVELIN is constructed in linear units, which renders the scale factors in the parameter sets unambiguous physical meanings. This feature is the ultimate reason why we can constrain the program using the (linear) flux ratio between the emission line and the continuum.

\section{Data Selection and Reduction}
\subsection{SDSS-RM Quasars}
These quasars make up a subset of SDSS-RM quasars used in \citet{Shen:2015fn}. The original dataset includes 32 epochs of BOSS spectra of 15 quasars (We will use the RMID in \citet{Shen:2015fn} to denote each of these quasars in the following). But we found that a significant fraction of the broad H$\beta$ line is also masked when masking the atmospheric lines in the spectra of the quasar RMID$=$769. We also found strong contamination from other broad emission lines in the line+continuum band of three other quasars (i.e. RMID$=$101, 229, 320). Thus these four quasars are excluded in the following analysis, which leaves the consequent sample size as 11. The observations have baselines of $\sim$180 days and the wavelength coverage is from $\sim$3650 \AA~ to 10400 \AA. The sampling is quite even, with a cadence of $\sim$5-6 days. After the pipeline processing the spectrophotometry, it has a nominal accuracy of $\sim$5\%.

Due to the difficulty in the direct absolute flux calibration induced by the use of a fiber spectrograph, the spectra were further calibrated using the code PrepSpec, which utilizes narrow emission lines (e.g. [OIII] $\lambda$4959, 5007) as flux calibrators. This procedure improves the calibration of the relative spectrophotometry to an accuracy of $\lesssim 2\%$. For further details about PrepSpec, see \citet{Shen:2014hk}. 

\subsection{SDSS Stripe 82 Quasars}

SDSS Stripe 82 (S82) is a dedicated multi-epoch survey which covers a relatively small region on the sky \citep{Abazajian:2009ef}, with $-50^{\circ} < \rm R.A. < 59^{\circ},\ \boldsymbol{-1^{\circ}.25 < \rm Decl. < 1^{\circ}.25}$. \citet{Hernitschek:2015iw} selected 362 targets with proper line+continuum and pure continuum bands from the 9258 S82 catalog quasars. Additional information about these quasars can be found in the Catalog of Quasar Properties from SDSS DR7 \citep{Shen:2011dg}. From this sample we further pick out the ones within the redshift range from 0.555 to 0.846 (275 quasars in total). Such a redshift range makes the broad Mg \textsc{ii}  emission line the indicator of the BLR sizes.

The original S82 quasar light curves are in unit of magnitude, so in order to utilize line-to-continuum ratio to constrain the calculation, we convert the magnitudes into fluxes (in arbitrary units). The potential caveat induced by the fact that in a DRW process, each value could be either positive or negative, while the physical fluxes are always positive, can be addressed by subtracting the mean fluxes during the observations, which was implemented in the original JAVELIN code.

\subsection{Line-to-Continuum Flux Ratio Calculation}

We calculate the line-to-continuum flux ratios of the selected quasars by fitting and integrating over the spectra to fix them when using JAVELIN to calculate the lags between broad emission lines and the continuum. For SDSS-RM quasars and those with repeated spectroscopic observations in S82 quasars, we use the RMS spectra in the calculation, and use single epoch spectra for other quasars in the S82 catalog.

We use a specialized spectra fitting IDL code to fit all the spectra. Before fitting we firstly mask out the points where the ANDMASK (set nonzero if the flux measurements at the corresponding wavelength are problematic at all epochs) value is nonzero, along with three windows where the contribution of sky lines is not negligible. We fit three Gaussian components for H$\alpha$, H$\beta$, and Mg \textsc{ii}  lines respectively, which are used as reverberation mapping target lines.

After the fitting we convolve the line and continuum flux spectra with the total transmission curve using Eq. (1) in \citet{Bessell:2012bq}

\begin{equation}
F_{\nu} = \frac{\int f_{\nu}\left(\nu\right)S_{x}\left(\nu\right)d\nu/\nu}{S_{x}\left(\nu\right)d\nu/\nu} = \frac{\int f_{\lambda}\left(\lambda\right)S_{x}\left(\lambda\right)\lambda d\lambda}{\int S_{x}\left(\lambda\right)cd\lambda},
\end{equation}
where $S_{x}$ is the total transmission curve in the filter band $x$ and $c$ is the speed of light. The output line and continuum flux give the ratio needed as input to the program.

\subsection{Conversion from Spectra to Light Curves}

For SDSS-RM quasars the original data are spectra. To convert them into multi-epoch photometric data we convolve them with the total transmission curve, similar to the procedure in the line-to-continuum flux ratio determination, and the errors are propagated to photometric errors.

\section{Results}

\subsection{SDSS-RM Quasars} 
\label{sub:sdss_rm_quasars}

\addtolength{\oddsidemargin}{-.2in}
\addtolength{\evensidemargin}{-.2in}
\addtolength{\textwidth}{0.4in}
\begin{figure}[htb!]
\begin{adjustwidth}{-5cm}{-5cm}
\begin{center}
\gridline{\fig{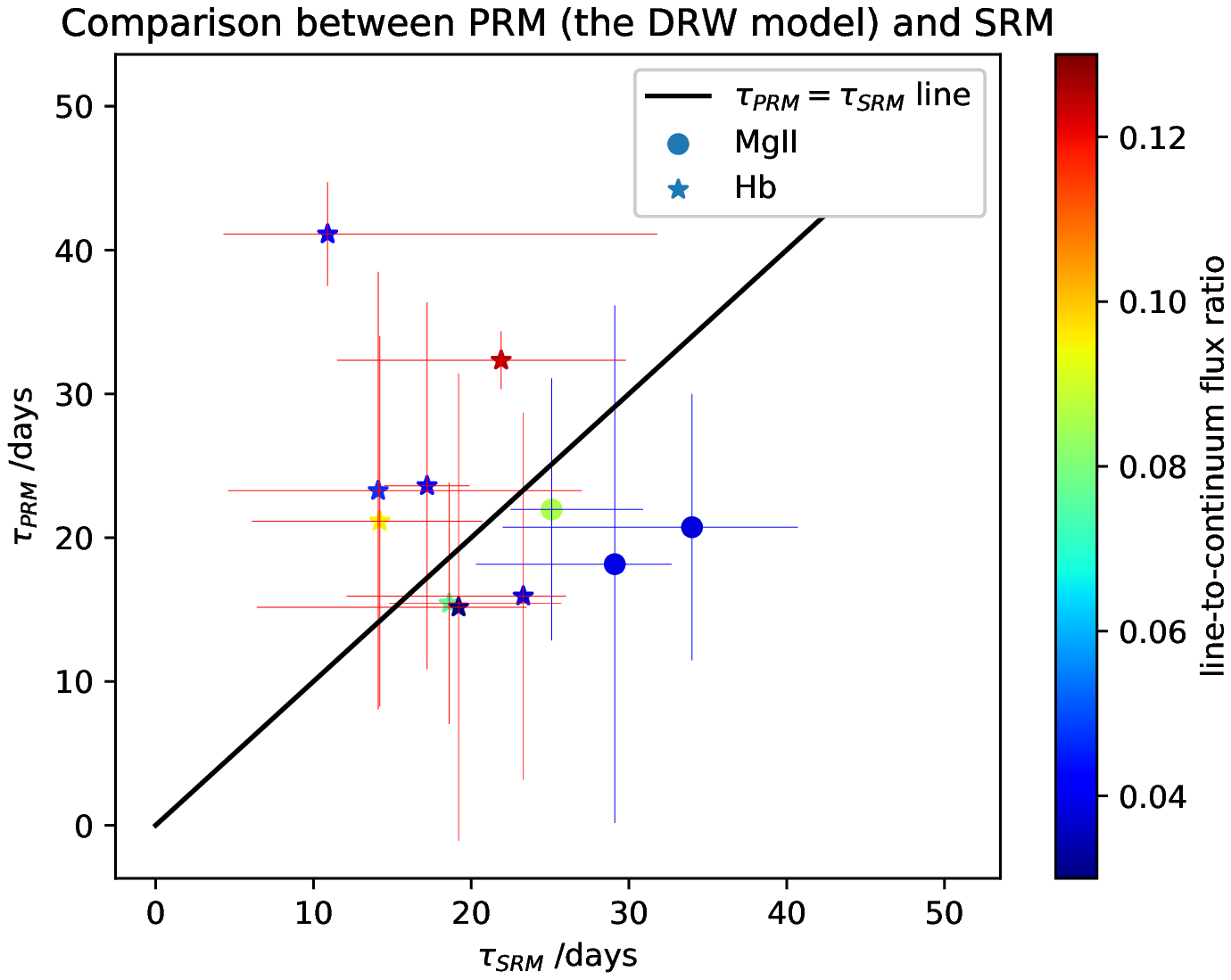}{0.6\textwidth}{}}
\gridline{\fig{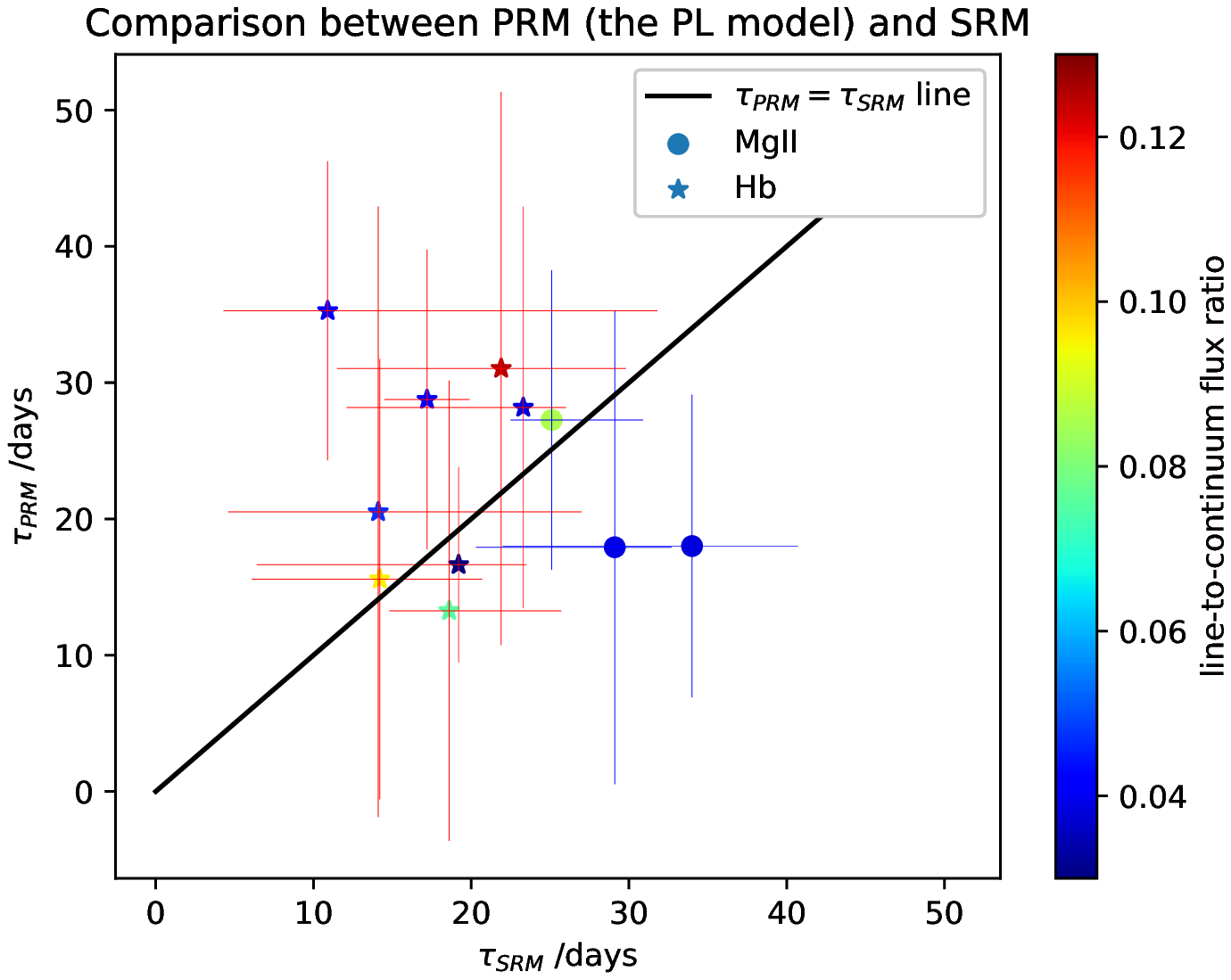}{0.6\textwidth}{}}
\end{center}
\end{adjustwidth}

\caption{Consistency between broadband PRM results and SRM results from \citet{Shen:2015fn}. Upper panel: the DRW model; Lower panel: the power-law model. Filled circles: Mg II lines; Stars: H$\beta$ lines. Black solid line shows $\tau_{PRM}=\tau_{SRM}$. \label{fig:f1}}
\end{figure}

\addtolength{\oddsidemargin}{.2in}
\addtolength{\evensidemargin}{.2in}
\addtolength{\textwidth}{-.4in}

We firstly show the broadband PRM results on the 11 SDSS-RM quasars. Fig. 1 shows the consistency between the SRM results and the broadband PRM results from this work. The upper panel shows the DRW model results and the lower panel shows the power-law model results. The lags calculated from different lines are marked by different labels, with filled circles for Mg \textsc{ii}  lines, and stars for H$\beta$ lines. The labels are color-coded by their line-to-continuum flux ratios. A significant fraction of the PRM results show good consistency with the SRM results, and we can see that the results under the two models are quite similar. The performance of the two models will be further discussed in section 5. But there is no clear indication on which broad emission line can give more consistent results with the SRM results. Table 1 lists the broadband PRM results and the information needed in JAVELIN run.

\begin{deluxetable*}{cccccccccccccc}[]
\tablecaption{Broadband PRM results of SDSS-RM quasars \label{tab:1}}
\tablenum{1}
\tablewidth{0pt}
 \setlength{\tabcolsep}{1.8pt}
\tablehead{
\colhead{RMID} & \colhead{Line} & \colhead{Redshift} & \colhead{Continuum} & \colhead{Line} & \colhead{$\tau_{SDSS-RM}$} & \colhead{log$L_{5100}$} & \colhead{$\tau_{R-L}$} & 
\colhead{$\tau_{cent,DRW}$} & \colhead{$\tau_{peak,DRW}$} &
\colhead{$\sigma_{\tau,DRW}$} & \colhead{$\tau_{cent,PL}$} & \colhead{$\tau_{peak,PL}$} & \colhead{$\sigma_{\tau,PL}$} \\
\colhead{(-)} & \colhead{(-)} & \colhead{(-)} & \colhead{band} &\colhead{band} &\colhead{(days)} & \colhead{(ergs$\cdot$s$^{-1})$} & \colhead{(days)} & 
\colhead{(days)} & \colhead{(days)} &
\colhead{(days)} & \colhead{(days)} & \colhead{(days)} & \colhead{(days)}
}
\startdata
191 & $\rm H\beta$ & 0.4418 & r & i & 23.3 & 43.646 & 21.793 & 24.3 & 17.1 & 14.1 & 20.0 & 21.1 & 21.9 \\
267 & $\rm H\beta$ & 0.5872 & r & i & 18.6 & 44.092 & 37.673 & 14.4 & 14.8 & 14.8 & 12.8 & 13.2 & 17.5 \\
272 & $\rm H\beta$ & 0.2628 & g & r & 21.9 & 43.929 & 30.483 & 32.2 & 31.6 & 2.7 & 40.3 & 37.2 & 12.7 \\
457 & Mg \textsc{ii}  & 0.6037 & r & g & 29.1 & 43.366 & 15.455 & 19.0 & 30.8 & 16.4 & 24.1 & 25.2 & 7.6 \\
589 &  Mg \textsc{ii}  & 0.7510 & r & g & 34.0 & 44.416 & 56.069 & 18.9 & 16.2 & 10.2 & 18.1 & 15.0 & 12.7 \\
645 & $\rm H\beta$ & 0.4738 & r & i & 14.2 & 44.109 & 38.468 & 18.7 & 19.6 & 12.6 & 13.4 & 8.9 & 18.7 \\
694 & $\rm H\beta$ & 0.5324 & r & i & 14.1 & 44.155 & 40.702 & 20.6 & 31.2 & 17.7 & 21.6 & 14.1 & 19.2 \\
767 &  Mg \textsc{ii}  & 0.5266 & r & g & 25.1 & 43.930 & 30.881 & 27.0 & 27.5 & 8.3 & 18.3 & 2.3 & 22.1 \\
775 & $\rm H\beta$ &  0.1725 & z & r & 19.2 & 43.541 & 19.158 & 13.0 & 12.6 & 14.6 & 14.2 & 16.2 & 8.7 \\
789 & $\rm H\beta$ & 0.4253 & g & i &17.2 & 43.685 & 22.861 & 27.1 & 20.6 & 11.6 & 23.3 & 31.4 & 17.5 \\
840 & $\rm H\beta$ & 0.2439 & r & i & 10.9 & 43.178 & 12.271 & 41.7 & 41.8 & 2.3 & 32.8 & 32.4 & 14.9 \\
\enddata
\end{deluxetable*}

\subsection{Stripe 82 Quasars} 
\label{sub:stripe_82_quasars}

We have also tested our code on the sparse and unevenly sampled light curves from SDSS Stripe 82. We chose the quasars with redshifts within the range from 0.555 to 0.846, and tried to determine the sizes of BLRs where the broad Mg \textsc{ii}  emission line originates. All of the samples are directly taken from \citet{Hernitschek:2015iw}. The upper panel of Fig. 2 shows the comparison between our results and the results from \citet{Hernitschek:2015iw}. We can clearly see that our yielded time lags cluster severely around the value of $\sim$400 days, which is characteristic of the bias induced by seasonal gaps in Stripe 82 light curves. With significant gaps in the light curves, the program tends to select time lag values which put the most of the points of the light curve in one band into the gaps of the other one, since mathematically it maximizes the likelihood. The reason why this phenomenon is significantly addressed in the results from \citet{Hernitschek:2015iw} is that they implemented a quite strong prior, imposing lags closer to the R-L relation larger prior probability in MCMC. The lower panel shows the comparison between our results and the estimations from R-L estimations. Still, the clustering around $\sim$400 days appears pronounced. It is notable that there seems to be a population for which our time lag estimation is rather consistent with the R-L relation values (those that are close to the black equality line), but we have to point out the fact that statistically the property of these quasars and their corresponding light curves show no sign of difference from those of the rest in the selected sample. 

\addtolength{\oddsidemargin}{-.2in}
\addtolength{\evensidemargin}{-.2in}
\addtolength{\textwidth}{0.4in}
\begin{figure}[htb!]
\begin{adjustwidth}{-5cm}{-5cm}
\begin{center}
\gridline{\fig{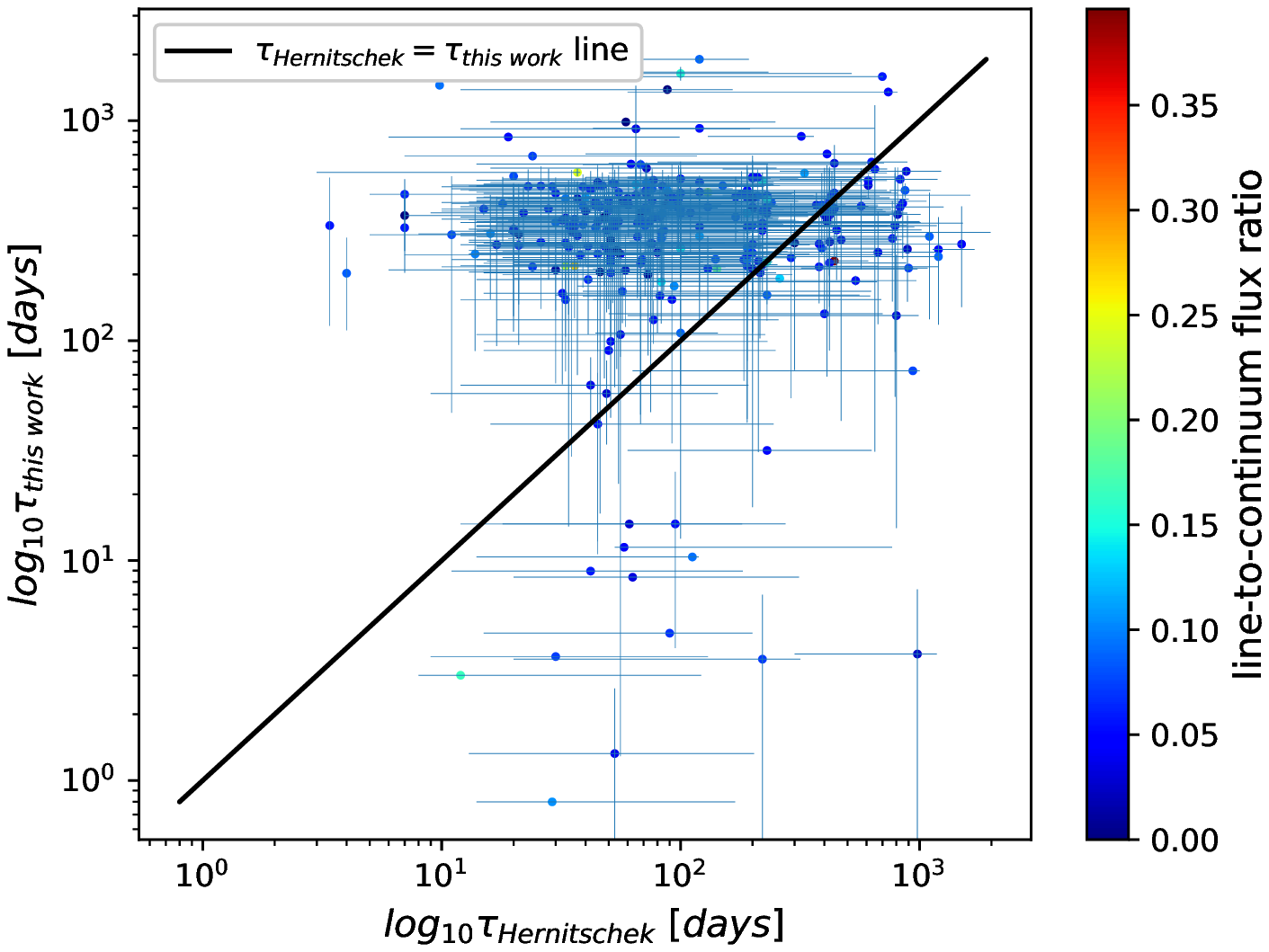}{0.65\textwidth}{}}
\gridline{\fig{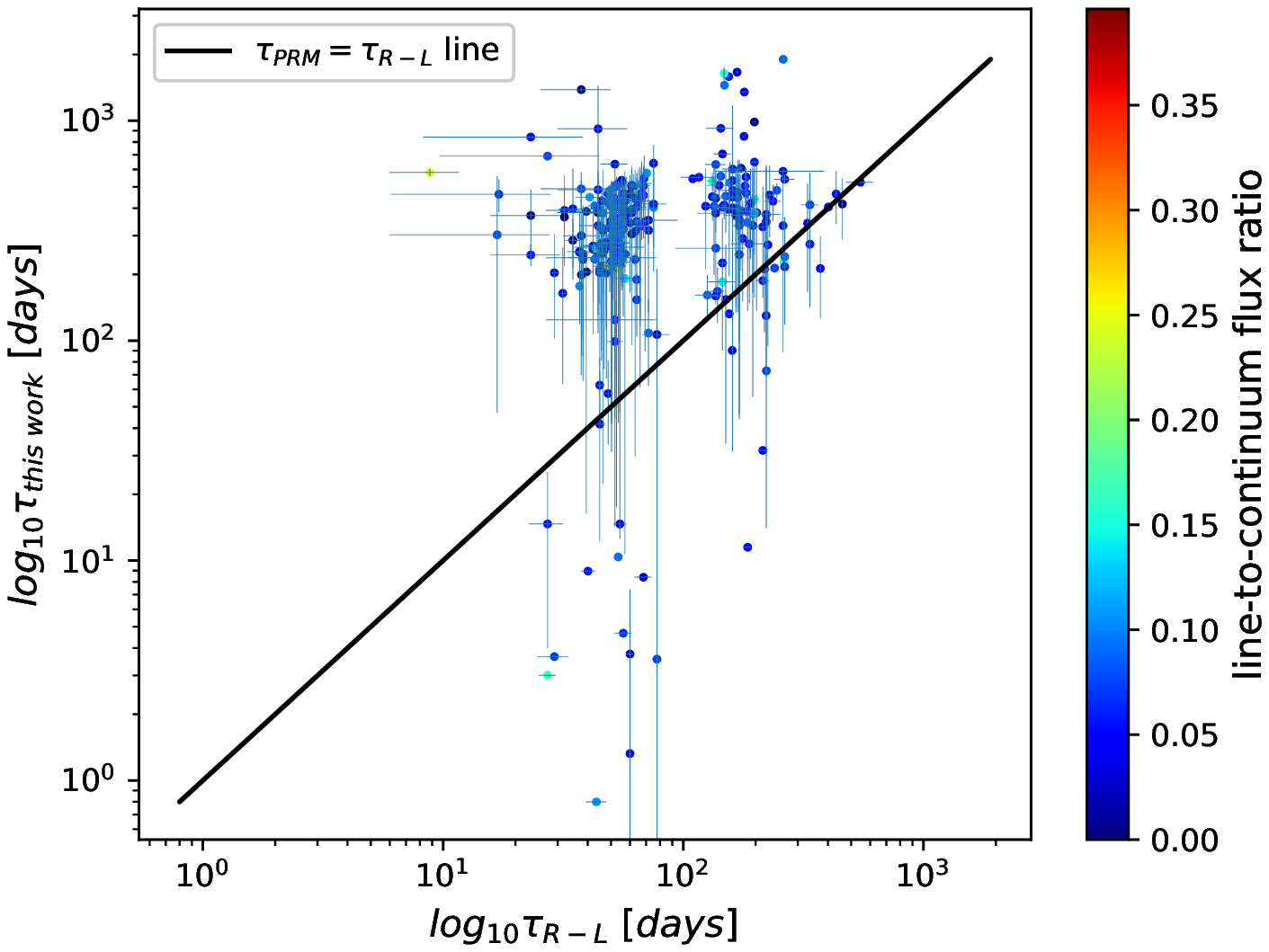}{0.65\textwidth}{}}
\end{center}
\end{adjustwidth}
\caption{Comparison of our results with the results from \citet{Hernitschek:2015iw} and the R-L relation estimations. Upper panel: Comparison with the results from \citet{Hernitschek:2015iw}; Lower panel: Comparison with the R-L relation. Black solid line shows the equality line.\label{fig:f2}}

\end{figure}
\addtolength{\oddsidemargin}{.2in}
\addtolength{\evensidemargin}{.2in}
\addtolength{\textwidth}{-.4in}

\section{Discussion} 
\label{sec:discussion}

\subsection{What Factor Leads to the Successful PRM lags?} 
\label{sub:what_factor_leads_to_successful_prm_detection_}

A natural question after we have carried out broadband PRM on the selected quasars is whether there exists some key factor(s) which may be significant in determining the lags between the broad emission lines and the continuum. Given the even and intensive sampling cadences for all of the quasars in this dataset, we note that the sampling quality isn't a major factor for SDSS-RM quasars.

After some investigations we found that there exists a critical line-to-continuum flux ratio, above which the mean value of the ratios between the broadband PRM lag obtained in this work and the corresponding SRM results from \citet{Shen:2015fn}, $\tau_{PRM}/\tau_{SRM}$, are closer to unity, and the scatter is significantly reduced. Fig. 2 shows the plot of $\tau_{PRM}/\tau_{SRM}$ versus the line-to-continuum flux ratios for the DRW and PL models. From this figure we estimate the critical flux ratio is about $6\%$ for DRW model, but due to the very limited sample size the estimation is by all means crude, and may be biased. The statistics of the lag ratios (i.e. $\tau_{PRM}/\tau_{SRM}$), separated by the critical flux ratio  are listed in Table 2.

\begin{deluxetable*}{ccccc}[b!]

\tablecaption{Broadband PRM results of SDSS-RM quasars \label{tab:2}}
\tablecolumns{11}
\tablenum{2}
\tablewidth{0pt}
\tablehead{
\colhead{Model} & \colhead{$[\frac{\tau_{PRM}}{\tau_{SRM}}]_{big}$\tablenotemark{a}} & \colhead{$\sigma[\frac{\tau_{PRM}}{\tau_{SRM}}]_{big}$\tablenotemark{b}} & \colhead{$[\frac{\tau_{PRM}}{\tau_{SRM}}]_{small}$\tablenotemark{c}} & \colhead{$\sigma[\frac{\tau_{PRM}}{\tau_{SRM}}]_{small}$\tablenotemark{d}}
}

\startdata
DRW & 1.17 & 0.32 & 1.36 & 1.05 \\
PL & 1.08 & 0.25 & 1.37 & 0.86 \\
\enddata

\tablenotetext{a}{mean value of the lag ratios with line-to-continuum flux ratios larger than the critical value}
\tablenotetext{b}{standard deviation of the lag ratios with line-to-continuum flux ratios larger than the critical value}
\tablenotetext{c}{The same as $a$, but of those with flux ratios smaller than the critical value}
\tablenotetext{d}{The same as $b$, but of those with flux ratios smaller than the critical value}

\end{deluxetable*}

\addtolength{\oddsidemargin}{-.2in}
\addtolength{\evensidemargin}{-.2in}
\addtolength{\textwidth}{0.4in}
\begin{figure}[htb!]
\begin{adjustwidth}{-5cm}{-5cm}
\begin{center}
\gridline{\fig{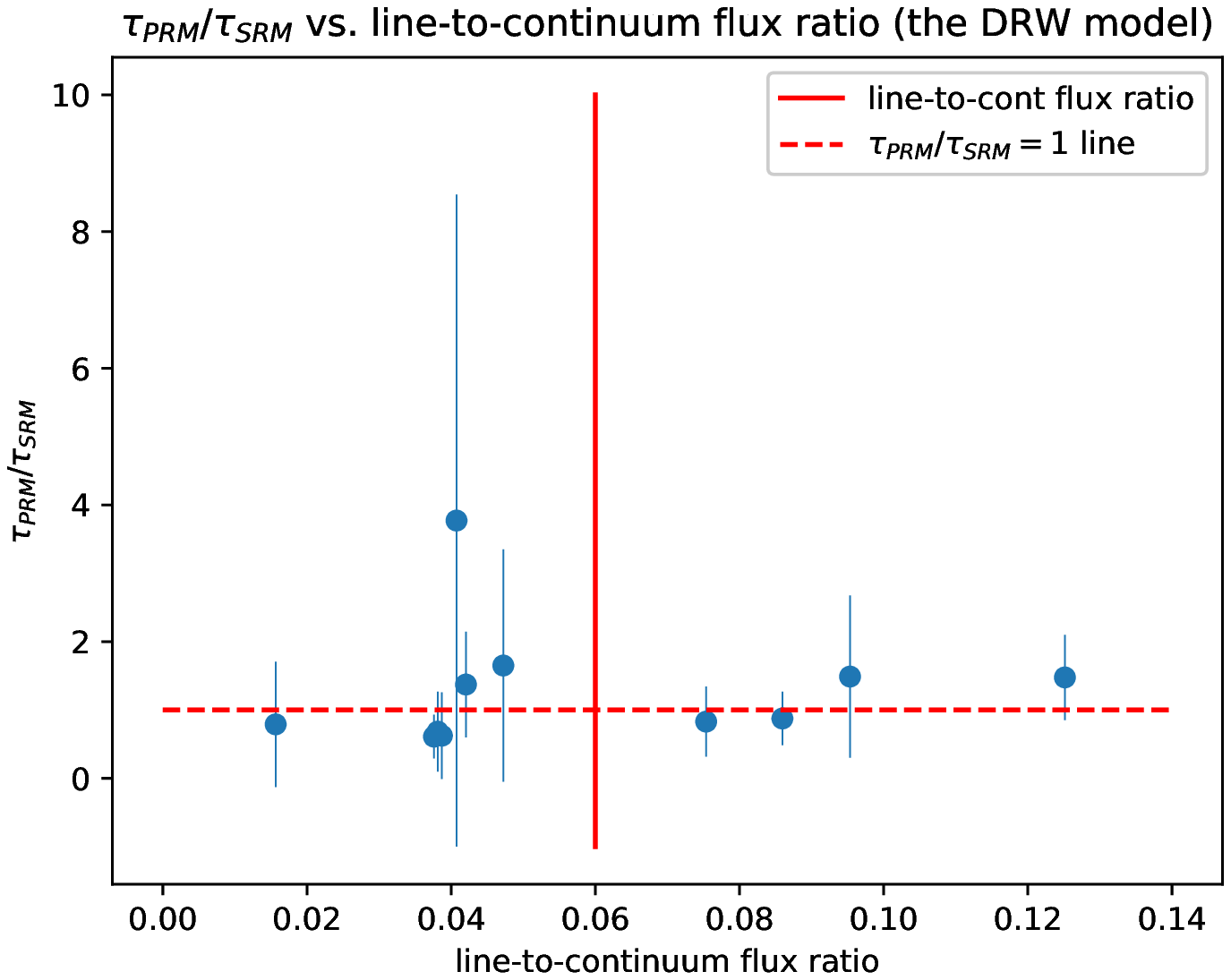}{0.65\textwidth}{}}
\gridline{\fig{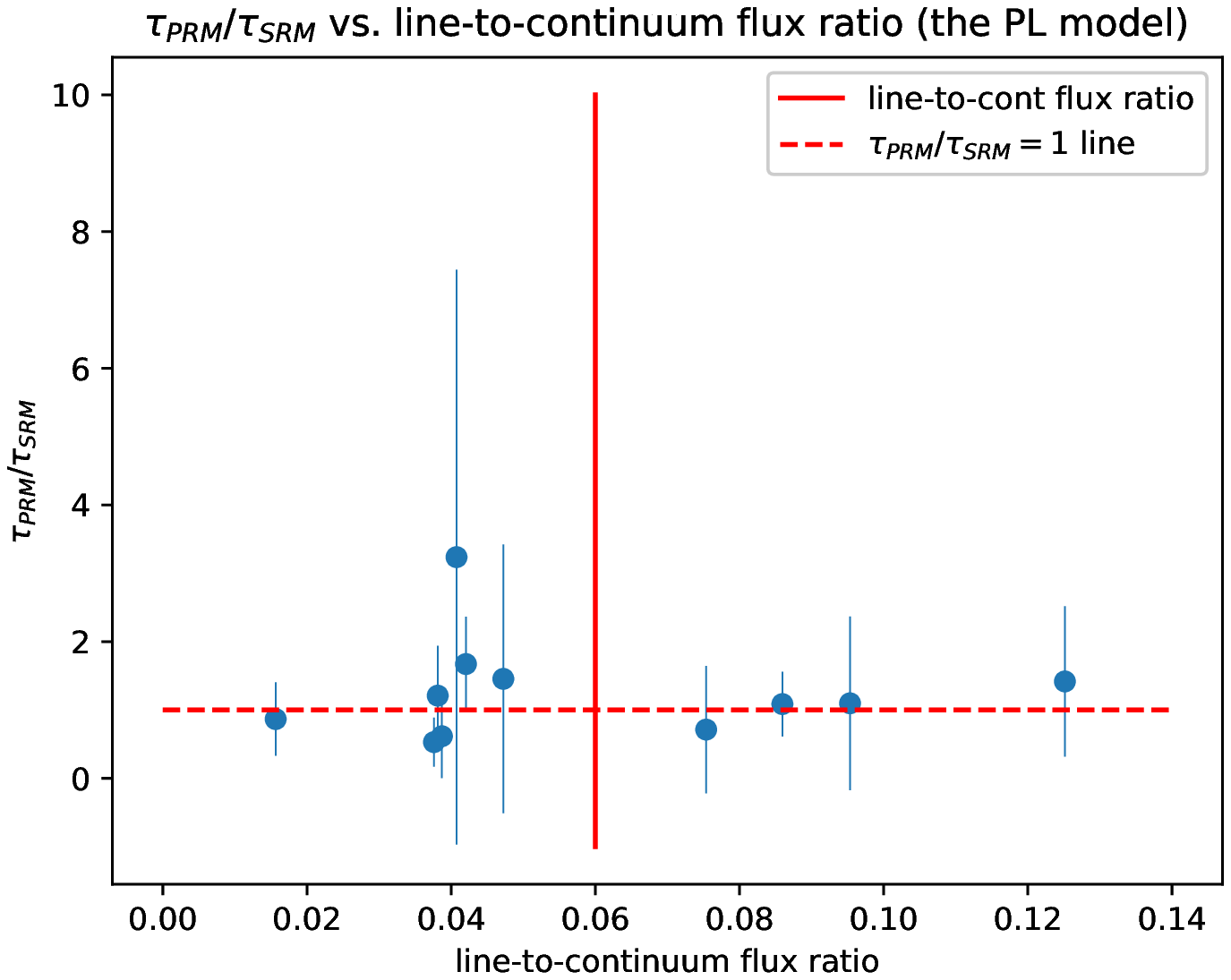}{0.65\textwidth}{}}
\end{center}
\end{adjustwidth}
\caption{$\tau_{PRM}$/$\tau_{RL}$ vs. line-to-continuum flux ratio plot. Upper panel: DRW model; Lower panel: Power-law model. The red vertical solid line marks the critical flux ratio value under each model. The dashed horizontal line shows $\tau_{PRM}/ \tau_{RL}=1$\label{fig:f3}}
\end{figure}
\addtolength{\oddsidemargin}{.2in}
\addtolength{\evensidemargin}{.2in}
\addtolength{\textwidth}{-.4in}

\subsection{Which Performs Better, DRW or Power-Law Model?} 
\label{sub:which_performs_better_drw_or_power_law_}

The power-law model has been widely utilized in structure function analysis of quasar optical variability 
\citep[e.g.,][]{2010ApJ...714.1194S,Anonymous:NTBTE_Lk,simkoz:2017vi,Anonymous:DtBp8yhK}. \citet{Hernitschek:2015iw} found that the power-law model performs better when describing light curves of similar sampling qualities to SDSS S82 data, which are highly unevenly and sparsely sampled. A major part of our modification to the original JAVELIN code is the addition of such a power-law model, and as is shown in Fig. 1, the power-law model shows similar results when used to estimate the BLR time lags as DRW did. In this subsection we try to find out whether this similarity is caused by random factors by doing repeated simulations of broadband PRM procedures.

Real photometric data of a certain quasar are only specific realizations of the underlying DRW process. In the simulation, we generate more theoretical light curves of both the pure continuum band and line+continuum band, using the lags from \citet{Shen:2015fn}. All of the simulated light curves possess sampling intervals similar to those of SDSS-RM. Due to the relatively short length of the light curves, JAVELIN code cannot constrain the DRW or power-law model parameters very well, so we use $\sigma = 0.1$ (The unit is the same as the flux unit in the output spectra of PrepSpec code) , $\tau = 5\ \rm{days}$, and a top-hat transfer function with a width of 5 days for all the simulated light curves. After generating the light curves we use them as input to JAVELIN and calculate the lags in an identical way as we treated the real data, except that we only calculate once for each simulated realization (recall that in the calculation using real data, we calculated 200 times for each quasar). We calculated the lags for 200 simulated realizations for each quasar under DRW and power-law model, respectively, and compare the distributions of the lags under the two models with the real lag values used in light curve generation. The result of the quasar RMID$=$191 is shown as an example in Fig. 4. The blue histogram is the distribution of the lag values which maximizes the posterior probability in each of the 200 MCMC runs under the DRW model, the green one is the same, but for the PL model. The red one is the distribution of the lags which are the median value in each of the MCMC runs under the PL model. The vertical line denotes the SRM result from \citet{Shen:2015fn}. It is obvious that the distribution of lags show a more pronounced peak and the peak is much closer to the SRM result under DRW model. Given the relatively large simulated sample size in this test, we are more convinced that under the sampling conditions of SDSS-RM project and of course, LSST \citep{Ivezic:2008ub}, the DRW model is indeed a better description of the quasar optical variability.

Given the argument in \citet{Hernitschek:2015iw} that the power-law model performs better under the sparsely sampled galaxies, we also generated light curves with typical sampling patterns of S82 ones, and carried out the same tests as above. Again, for all the objects, the DRW outperforms the power-law model. See Fig. 5 for an example. Note that this seemingly contradiction between the two conclusions could be due to the differences in our lag calculation program, e.g. \citet{Hernitschek:2015iw} used a piror distribution which prefers lag values  closer to R-L relation while we did not, we ran 200 times of MCMC to determine the time lags for each realization of the quasar variability (i.e. light curve), and we fixed the line-to-flux continuum ratio whenever available (i.e. whenever there are good spectra for a certain quasar to calculate this ratio) while they did not.


\begin{figure}[htb!]
\plotone{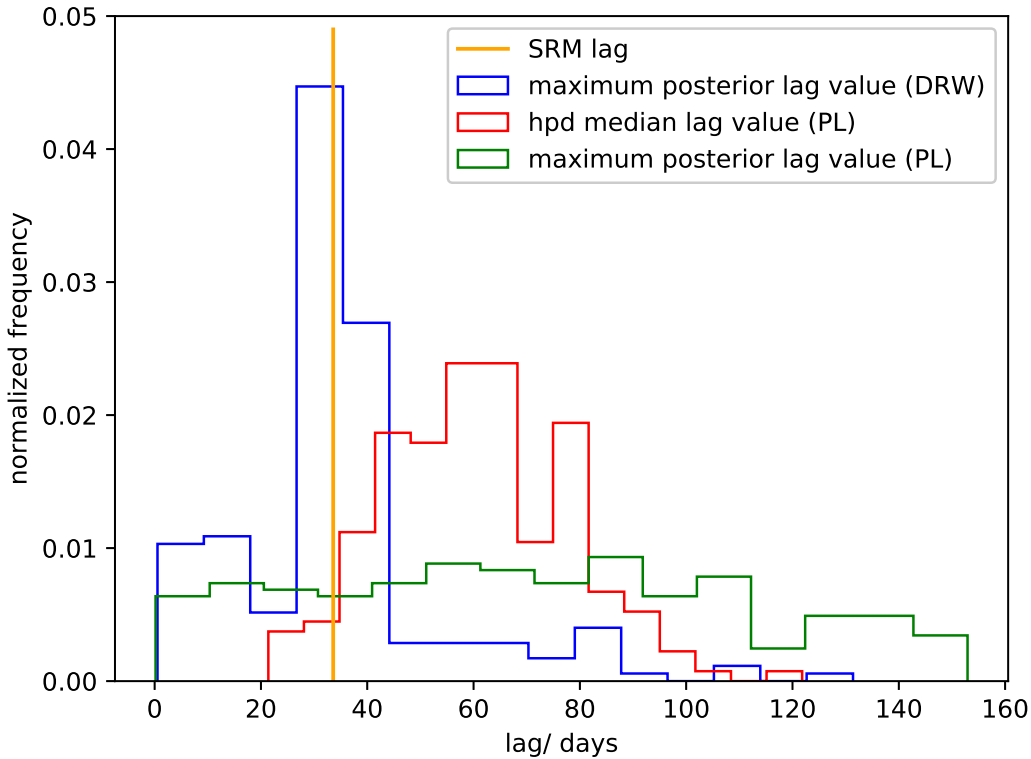}
\caption{Comparison between the distribution of lag results under DRW and power-law models in the simulation (simulated SDSS-RM cadence light curves) for the source RMID=191. Blue histogram: distribution of maximum-posterior-value lags under DRW model; Green histogram: distribution of maximum-posterior-value lags under power-law model; Red histogram: distribution of median lag value in each MCMC run under power-law model. Vertical line: time lag estimation from \citet{Shen:2015fn}. \label{fig:f4}}
\end{figure}

\begin{figure}[htb!]
\plotone{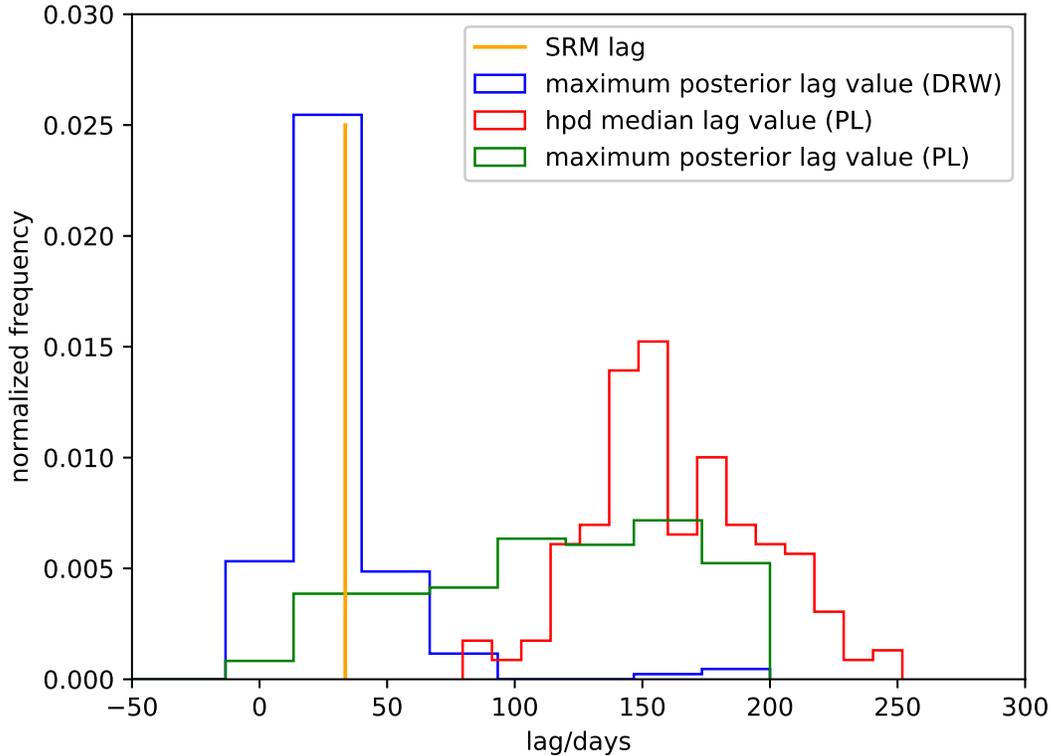}
\caption{Same as Fig. 4 but for the simulated S82 cadence light curves.\label{fig:f5}}
\end{figure}


\subsection{What Do We Learn from the Results of Stripe 82 Lags?} 
\label{sub:what_do_we_learn_from_the_results_of_stripe_82_lags_}

Clearly our program doesn't work as well on the sparse and unevenly light curves from S82 as \citet{Hernitschek:2015iw} did.  Many estimations are affected by the seasonal observation gaps and thus the lags cluster around the value of $\sim$400 days. We could have yielded better results by imposing similar priors to \citet{Hernitschek:2015iw} to make the results closer to the R-L relation, but given the empirical nature of this relation, we chose not to implement this constraint in order to avoid possible bias for quasars which deviate non-negligibly from the relation. Based on this philosophy the best solution we can give in order to get rid of the influence of the observational gap -- if only sparse and uneven sampled data are available -- is to ignore the lag values with apparent clustering.

On the other hand, from these results one should be educated that pure broadband PRM should be carried out mainly (if not exclusively) on the data of similar sampling quality to the SDSS-RM quasars. Fortunately LSST will be able to present this particular bonanza of quasar variability in the next a few years, and we will then be able to get rid of the influence of the observational gaps in principle.



\section{summary} 
\label{sec:summary}

We extended and modified the broadband photometric reverberation mapping code JAVELIN, which was made by \citet{2010ascl.soft10007Z} by adding the power-law structure function model and modifying the priors imposed along with the way to get an estimation of the time lags between the broad emission lines and the underlying continuum from the results of JAVELIN. By testing the algorithm on the selected 11 SDSS-RM project quasars \citep{Shen:2015fn}, whose light curves are produced by convolving multi-epoch spectra with transmission curves,  and also 275 SDSS Stripe 82 quasars, we found:

\begin{itemize}
	\item[(1)] Fixing the line-to-continuum flux ratio at the value determined from spectra fitting helps much in constraining the yielded lags within the physically realistic range;
	\item[(2)] There is a critical line-to-continuum flux ratio value, above which the mean ratio of the PRM lags to the corresponding SRM results from \citet{Shen:2015fn} is closer to one, and the scatter is significantly reduced by a factor of $\sim$3. From simple visual inspection of Fig. 3, we estimate this critical value as $\sim 6\%$, but it's a crude estimation and may be biased due to the very limited sample size.
	\item[(3)] Although  the results of DRW and PL models are quite similar for the SDSS-RM quasars, further PRM simulations on mock data (whether they are of similar qualities to SDSS-RM or Stripe 82 quasar light curves) show that DRW model performs better than PL model.
	\item[(4)] For the Stripe 82 quasars, our programs tend to give biased estimations on the time lags when the prior selecting consistent values with R-L relation is removed. This is principally due to the existence of seasonal observational gaps. We argue that the clustering around biased values will vanish given the data with little gaps and yield better estimations, which is highly expected in the era of LSST.
\end{itemize}

We thank the anonymous referee for elaborate review and helpful comments on the manuscript, and Yue Shen for providing us the SDSS-RM spectra. The work is partically supported by the Ministry of Science and Technology of China under grant 2016YFA0400703, the NSFC grant No.11373008 and No.11533001, and the National Key Basic Research Program of China 2014CB845700. 

\appendix
\section{Covariance Matrix Entries under Power-law Model}

When calculating the covariance matrices we assume the transfer function is of a top-hat form:

\begin{equation}
	\Psi\left(t - \tau\right) = \frac{s_{line}}{t_2-t_1}\ \ \rm{for}\ \emph t_1 \le \emph t - \tau \le \emph t_2,
\end{equation}
where $s_{line}$ is the amplitude of the transfer function, i.e. the line-to-continuum flux ratio.

\subsection{The Covariance Matrix between the Continuum and One Line}
\begin{equation}
	\langle f_c\left(t_i\right), f_l\left(t_j\right) \rangle = \frac{s_{line}A^2}{2\left(t_2 - t_1\right)\left(\gamma + 1\right)} \left(-|t_{hig}|^{\gamma} t_{hig} + |t_{low}|^{\gamma}t_{low} + 2\left(\gamma+1\right)\left(t_{hig} - t_{low}\right)t_{obs}^{\gamma}\right),
\end{equation}
where $t_{hig} = t_{i} - t_j - t_2$, $t_{low} = t_i - t_j - t_1$ and $A$ is the amplitude of the structure function.


\bibliography{PRM}
\bibliographystyle{aasjournal}

\end{document}